# Head-on collision of ion acoustic shock waves in collisionless plasma with pair-ions and electrons

M S Alam[1,3], M R Talukder[2*]

[1] Department of Applied Mathematics, University of Rajshahi, Rajshahi-6205, Bangladesh

[2] Plasma Science and Technology Lab, Department of Applied Physics and Electronic Engineering, University of Rajshahi, Rajshahi-6205, Bangladesh

[3] Department of Mathematics, Chittagong University of Engineering and Technology, Chittagong-4349, Bangladesh

[*] Corresponding author: mrtalukder@ru.ac.bd

**Abstract**

Unmagnetized collisionless plasma system consisting of positive- and negative ions, and electrons are considered to study the head-on collision of ion-acoustic shock waves (IASWs) and its effects on the formation of shock (monotonic and oscillatory) waves and phase shift. The soliton solution is derived from the two-sided Korteweg-de Vries Burger (KdVB) equations. The KdVB equations are obtained using extend Poincaré-Lighthill-Kuo (ePLK) method. It is assumed that the negative ions are immobile and the electrons follow the Boltzmann energy distribution in the plasma. The effects of plasma parameters such as density ratios and kinematic viscosities on electrostatic shock profiles, phase shift, amplitudes, formation of shock (monotonic and oscillatory) as well as on the soliton solution are investigated. It is found that the density ratio of negative to positive ions plays a vital role on the formation of shock waves and phase shift after collision.

**1. Introduction**

The non-linear propagation of ion-acoustic waves is one of the most studied aspects in plasma physics. It is not always an easy task to obtain direct solution of the hydrodynamic-like model equations which govern their evolution. It is obvious that the asymptotic technique is both simpler and more informative to obtain the solution of such evolution equations. In 1966 Washimi and Taniuti [1] used the reductive perturbation technique to derive Korteweg-de Vries (KdV) equation to study the wave propagation in plasma systems consisting of cold ions and hot electrons. Later on, the asymptotic technique is extensively used for making the complicated systems of partial differential equations (PDEs) more tractable model equations that describe the wave propagation when the combined effects of nonlinear steepening and dispersion balance each other to give rise to the formation of localized structures [2,3]. On the other hand, recently many authors are attracted to investigate the interactions between waves due to their potentiality with the extended Poincaré-Lighthill-Kuo (ePLK) method [4-6].

Negative ion plasmas have attracted great interest to the researchers due to the wide technological applications such as in neutral beam source [7], plasma processing reactor [18], semiconductor and material processing [8], and so on. The presence of negative ions in addition to electrons and positive ions in plasmas are ascribed by [9-12] and noted that the negative ions significantly modify the characteristic of plasma phenomena. These types of plasmas are observed in space and astrophysical objects and it can be produced in laboratory [13-27] as well. Von Goeler et al [20] have directed a beam of CsCl onto the hot tungsten plate of a Q-machine that forms plasma consisting of $Cs^+$, $Cl^-$ and electrons. Wong et al. [21] have studied experimentally the ion acoustic (IA) waves (IAWs) in negative ion plasmas with $SF_6^-$ species. The theoretical investigation on the properties of IAWs is reported by Angelo et al. [24], Angelo [25], and Galvez and Gary [26]. Song et al. [28] have also reported an experimental investigation of IAWs in Q-machine plasmas consisting of $K^+$, $SF_6^-$ and electrons. They observed that the phase velocity of ion-acoustic "fast" mode [9, 21] increases with increasing density ratio of negative to positive ions. The fast wave may form a negative-potential solitary waves [29, 30], when the nonlinear steepening is consider, in addition to dissipation it will either disperse, or more significantly form a shock [31]. Due to its importance, the shock waves are investigated theoretically and experimentally in space, astrophysical and laboratory plasmas [32-35]. The productions of shocks in Q-machine with negative ions are reported [36, 37]. Adak et al [38,39] have studied the magnetosonic shocks and investigated the effect of ion-ion interactions on the dynamics of nonlinear IAWs in fully collisional pair-ion ($C_{60}^+$, $C_{60}^-$) plasmas along with KdV Burger equation using reductive perturbation method and found that the positive (negative) potential IAWs exists when $T_- > T_+$ ($T_- < T_+$). Hossain et al [40] have investigated the nonlinear ion-acoustic monotonic as well oscillatory shock waves in negative ion plasmas with nonextensive electrons. Saeed and Mushtaq [41] have investigated the linear and nonlinear properties of low frequency IAWs in pair ion plasmas with Boltzmann distributed electrons along with Kadomtsev-Petviashvili (KP) equation and found that the structure of the IAWs are significantly affected by the ratio of electron number density and temperature. The collision of counter propagating solitary waves are studied theoretically [42,43] and experimentally [44] and found that both solitons lag in phase but the smaller soliton lags more than that of larger one. To the best of our knowledge no one studies the head-on collision of IA solitary and shock waves in a plasma systems consisting of positive ions ($K^+$), negative ions ($SF_6^-$) with Boltzmann energy distributed electrons. This report is organized as follows: Assumptions and hydrodynamic fluid equations are represented in section 2. Derivation of evolution equations and phase shifts with analytical solution are provided in section 3. Results and discussion are displayed in section 4. Conclusion is drawn in section 5.

## 2. Assumsions and hydrodynamic fluid equations

Unmagnetized collisionless three-component plasma system consisting of positive and negative ions with Boltzmann energy distributed electrons is considered. The normalized hydrodynamic fluid equations and Poisson's equation for the concerned plasma system are, respectively, given by

$$\frac{\partial n_{1i}}{\partial t} + \frac{\partial (n_{1i} v_{1i})}{\partial x} = 0, \qquad (1)$$

$$\frac{\partial n_{2i}}{\partial t} + \frac{\partial (n_{2i} v_{2i})}{\partial x} = 0, \qquad (2)$$

$$\frac{\partial v_{1i}}{\partial t} + v_{1i} \frac{\partial v_{1i}}{\partial x} + \frac{\partial \phi}{\partial x} + \eta_{1i} \frac{\partial^2 v_{1i}}{\partial x^2} = 0, \qquad (3)$$

$$\frac{\partial v_{2i}}{\partial t} + v_{2i} \frac{\partial v_{2i}}{\partial x} - m_{12i} \frac{\partial \phi}{\partial x} + \frac{T_{2ie}}{n_{2i}} \frac{\partial n_{2i}}{\partial x} + \eta_{2i} \frac{\partial^2 v_{2i}}{\partial x^2} = 0, \qquad (4)$$

$$\frac{\partial^2 \phi}{\partial x^2} = n_{1i} - n_{2i} - n_e, \qquad (5)$$

where, $n_{1i}$, $n_{2i}$, $n_e$, $v_{1i}$, $v_{2i}$ and $\phi$ are the densities of positive and negative ions, electrons, velocities of positive and negative ions, and electrostatic potential, respectively. $n_{1i0}$, $n_{2i0}$ and $n_{e0}$ are the unperturbed densities of positive and negative ions, and electrons, respectively. The charge neutrality condition is considered as $n_{1i0} = n_{2i0} + n_{e0}$. It can be converted to $n_{21i} + n_{e1i} = 1$, $n_{21i} = n_{2i0}/n_{1i0}$, $n_{e1i} = n_{e0}/n_{1i0}$, $m_{12i} = m_{1i}/m_{2i}$ and $T_{e2i} = T_e/(1 - n_{21i})T_{2i}$. $m_{1i}$ and $m_{2i}$ are the masses of positive and negative ions, $T_{1i}$ and $T_{2i}$ are the temperatures of positive and negative ions, and $T_e$ is the temperature of electrons, respectively. The densities $n_{2i}$ and $n_e$ are normalized by $n_{1i0}$, $\phi$ is normalized by $k_B T_e/e$, space coordinate is normalized by the electron Debye length $\lambda_{De} = \sqrt{k_B T_e/4\pi e^2 n_{e0}}$, ion velocity is normalized by the positive ion acoustic speed $c_s = \sqrt{k_B T_e/m_{1i}} \sqrt{(n_{e1i} + n_{21i})/n_{e1i}(1 - n_{21i})}$ and time coordinate is normalized by the positive ion plasma frequency $\omega_{p1i}^{-1} = \lambda_{De}/c_s$. Where $\sqrt{k_B T_e/m_{1i}}$, the positive ion acoustic speed in the absence of negative ions and $e$ is the electronic charge and $k_B$ is the Boltzmann constant. $\eta_{1i}$ and $\eta_{2i}$ are the normalized kinematic viscosities of positive and negative ions, respectively, obtained introducing in equations (3) and (4), where $\eta_{1i} = \eta_{d1i}\omega_{p1i}/m_{1i}n_{1i}c_s^2$, $\eta_{2i} = \eta_{d2i}\omega_{p2i}/m_{2i}n_{2i}c_s^2$. $\eta_{d1i}$ and $\eta_{d2i}$ are the dynamic viscosities of the positive and negative ions, respectively. It is assumed that $T_e \geq T_{1i} \gg T_{2i}$ [37]. Due to the effects of negative ions the electrons shielding is reduced and this effect demonstrates for the propagation and damping of linear IAWs in collisionless negative plasma with $SF_6^-$, which was observed in single-ended Q-machine plasmas [12,28]. The ion Landau damping is stronger for IAWs as obtained theoretically [45] and experimentally in a Q-machine plasma of $T_e \approx T_{1i}$, which has a big ballistic contribution to the wave propagation. In the single-ended Q-machine plasma of $T_e \geq T_{1i}$ with ion flow [12], the ion Landau damping is strong for IAWs but the ballistic effect can be neglected [46,47]. In the presence of negative ions, the factor $T_e \approx T_e/(1 - n_{21i})$ [37] is responsible for the increase of phase velocity, as a result the Landau damping of IAWs decreases, which is equivalent to the increase of $T_e/T_{1i}$ causes electron heating through electron cyclotron resonance [48]. The mass of negative ion ($SF_6^-$) (mass number: 146) is much larger than the positive ion mass ($K^+$) (mass number: 39). Thus one can assume that the negative ion is immobile and ion-acoustic mode is a positive ion acoustic (PIA) mode. Thus, no loss of generality is assumed that $T_{1i} \approx 0$ instead of $T_e/T_{1i} \geq 1$ because in presence of negative ions $T_e$ is replaced by $T_e/(1 - n_{21i})$.

**3. Derivation of evolution equations and analytical solution**

It is assumed that there are two solitons $S_L$ (propagates toward left) and $S_R$ (propagates toward right) in the considered plasma system approach toward each other. Initially ($t \to \pm\infty$) they are asymptotically far apart and propagate toward each other. After some time they interact and collide ($t = 0$) and then depart. As the collisions of soliton are elastic, the amplitudes of solitons will remain unchanged due to the interaction between them but after the collision each soliton acquires an additional phase shift. There are two ways of interaction between the two solitons in one dimensional system. One is head-on collision, in which the angle between the two propagating solitons is equal to $\pi$ and other is overtaking collision, in which the angle between the two propagating solitons is equal to 0. The phenomena pertaining to overtaking collision is beyond the scope of this study. To study the effects of head-on collisions on IA solitary waves (IASWs), the electrostatic resonance and their corresponding phase shifts can be derived from the two-sided Korteweg-de Varies Burger ( KdVB) equations considering the plasma system composing of unmagnetized collisionless positive ions and negative ions with Boltzmann energy distributed electrons using ePLK method. With respect to the ePLK method the stretched variables can be expanded as

$$\left.\begin{array}{l} \xi = \varepsilon(x - v_p t) + \varepsilon^2 \varphi_{10}(\zeta,\tau) + \varepsilon^3 \varphi_{11}(\xi,\zeta,\tau) + \cdots\cdots \\ \zeta = \varepsilon(x + v_p t) + \varepsilon^2 \varphi_{20}(\xi,\tau) + \varepsilon^3 \varphi_{21}(\xi,\zeta,\tau) + \cdots\cdots \\ \tau = \varepsilon^3 t \end{array}\right\}, \quad (6)$$

and the depended variables can be expanded around the equilibrium values in powers of $\varepsilon$ as

$$\left.\begin{array}{l} n_{1i} = 1 + \varepsilon^2 n_{1i}^{(1)} + \varepsilon^3 n_{1i}^{(2)} + \varepsilon^4 n_{1i}^{(3)} + \cdots\cdots \\ n_{2i} = n_{21i} + \varepsilon^2 n_{2i}^{(1)} + \varepsilon^3 n_{2i}^{(2)} + \varepsilon^4 n_{2i}^{(3)} + \cdots\cdots \\ v_{1i} = \varepsilon^2 v_{1i}^{(1)} + \varepsilon^3 v_{1i}^{(2)} + \varepsilon^4 v_{1i}^{(3)} + \cdots\cdots \\ v_{2i} = \varepsilon^2 v_{2i}^{(1)} + \varepsilon^3 v_{2i}^{(2)} + \varepsilon^4 v_{2i}^{(3)} + \cdots\cdots \\ \phi = \varepsilon^2 \phi^{(1)} + \varepsilon^3 \phi^{(2)} + \varepsilon^4 \phi^{(3)} + \cdots\cdots \\ n_e = n_{e1i} e^{\phi T_{2ie}} \end{array}\right\}, \quad (7)$$

where $v_p$ is the phase velocity of IASWs. From Eq. (1) the following operators are obtained

$$\left.\begin{array}{l} \partial_t = \varepsilon^3 \partial_\tau - v_p \varepsilon(\partial_\xi - \partial_\zeta) + v_p \varepsilon^3 (\partial_\zeta \varphi_{10} \partial_\xi - \partial_\xi \varphi_{20} \partial_\zeta) + \cdots\cdots \\ \partial_x = \varepsilon(\partial_\xi + \partial_\zeta) + \varepsilon^3 (\partial_\zeta \varphi_{10} \partial_\xi + \partial_\xi \varphi_{20} \partial_\zeta) + \cdots\cdots \end{array}\right\}. \quad (8)$$

As $\eta_{1i}$ and $\eta_{2i}$ in Eqs. (3) and (4) are small, therefore one may consider $\eta_{1i} = \varepsilon\eta_1$ and $\eta_{2i} = \varepsilon\eta_2$. Inserting Eqs. (6)-(8) into the Eqs. (1)-(5) one can obtain a set of partial differential equations (PDEs) in terms of $\varepsilon$. Now collecting the terms of equal powers of $\varepsilon$, the smallest order of $\varepsilon$ yields

$$-v_p \partial_\xi n_{1i}^{(1)} + v_p \partial_\zeta n_{1i}^{(1)} + \partial_\xi v_{1i}^{(1)} + \partial_\zeta v_{1i}^{(1)} = 0, \quad (9)$$

$$-v_p \partial_\xi v_{1i}^{(1)} + v_p \partial_\zeta v_{1i}^{(1)} + \partial_\xi \phi^{(1)} + \partial_\zeta \phi^{(1)} = 0, \quad (10)$$

$$-v_p \partial_\xi n_{2i}^{(1)} + v_p \partial_\zeta n_{2i}^{(1)} + n_{21i} \partial_\xi v_{2i}^{(1)} + n_{21i} \partial_\zeta v_{2i}^{(1)} = 0, \quad (11)$$

$$-v_p \partial_\xi v_{2i}^{(1)} + v_p \partial_\zeta v_{2i}^{(1)} - m_{21i} \partial_\xi \phi^{(1)} - m_{21i} \partial_\zeta \phi^{(1)} = 0, \quad (12)$$

$$n_{1i}^{(1)} - n_{2i}^{(1)} - n_{e1i} T_{e2i} \phi^{(1)} = 0. \quad 13)$$

Solving Eqs. (9)-(13) one can determine the following relations consisting of different plasma parameters as

$$\left.\begin{array}{l}\phi^{(1)} = \phi_\xi^{(1)} + \phi_\zeta^{(1)}, \ n_{1i}^{(1)} = 1/v_p^2\left[\phi_\xi^{(1)} + \phi_\zeta^{(1)}\right], \ v_{1i}^{(1)} = 1/v_p\left[\phi_\xi^{(1)} - \phi_\zeta^{(1)}\right]\\ n_{2i}^{(1)} = -n_{21i}m_{21i}/v_p^2\left[\phi_\xi^{(1)} + \phi_\zeta^{(1)}\right], v_{2i}^{(1)} = -m_{21i}/v_p\left[\phi_\xi^{(1)} - \phi_\zeta^{(1)}\right]\end{array}\right\}, \quad (14)$$

and the phase velocity is obtained as $v_p = [(1 + n_{21i}m_{21i})/n_{e1i}T_{e2i}]^{1/2}$. Where $\phi_\xi^{(1)}(\xi,\tau) \approx \phi_\xi^{(1)}$ and $\phi_\zeta^{(1)}(\zeta,\tau) \approx \phi_\zeta^{(1)}$ detect the right and left propagating waves, respectively. Taking the next higher order of $\varepsilon$ one can obtain the following relations:

$$\left.\begin{array}{l}\phi^{(2)} = \phi_\xi^{(2)} + \phi_\zeta^{(2)}, \ n_{1i}^{(2)} = 1/v_p^2\left[\phi_\xi^{(2)} + \phi_\zeta^{(2)}\right], \ v_{1i}^{(2)} = 1/v_p\left[\phi_\xi^{(2)} - \phi_\zeta^{(2)}\right]\\ n_{2i}^{(2)} = -n_{21i}m_{21i}/v_p^2\left[\phi_\xi^{(2)} + \phi_\zeta^{(2)}\right], v_{2i}^{(2)} = -m_{21i}/v_p\left[\phi_\xi^{(2)} - \phi_\zeta^{(2)}\right]\end{array}\right\}. \quad (15)$$

The next higher order of $\varepsilon$ provides

$$\partial_\tau n_{1i}^{(1)} - v_p \partial_\xi n_{1i}^{(3)} + v_p \partial_\zeta n_{1i}^{(3)} + \partial_\xi v_{1i}^{(3)} + \partial_\zeta v_{1i}^{(3)} + \partial_\xi\left(n_{1i}^{(1)} v_{1i}^{(1)}\right) + \partial_\zeta\left(n_{1i}^{(1)} v_{1i}^{(1)}\right)$$
$$+ v_p \partial_\zeta \varphi_{10} \partial_\xi n_{1i}^{(1)} - v_p \partial_\xi \varphi_{20} \partial_\zeta n_{1i}^{(1)} + \partial_\zeta \varphi_{10} \partial_\xi v_{1i}^{(1)} + \partial_\xi \varphi_{20} \partial_\zeta v_{1i}^{(1)} = 0 \quad (16)$$

$$\partial_\tau n_{2i}^{(1)} - v_p \partial_\xi n_{2i}^{(3)} + v_p \partial_\zeta n_{2i}^{(3)} + n_{21i} \partial_\xi v_{2i}^{(3)} + n_{21i} \partial_\zeta v_{2i}^{(3)} + \partial_\xi\left(n_{2i}^{(1)} v_{2i}^{(1)}\right) + \partial_\zeta\left(n_{2i}^{(1)} v_{2i}^{(1)}\right)$$
$$+ v_p \partial_\zeta \varphi_{10} \partial_\xi n_{2i}^{(1)} - v_p \partial_\xi \varphi_{20} \partial_\zeta n_{2i}^{(1)} + n_{21i} \partial_\zeta \varphi_{10} \partial_\xi v_{2i}^{(1)} + n_{21i} \partial_\xi \varphi_{20} \partial_\zeta v_{2i}^{(1)} = 0, \quad (17)$$

$$\partial_\tau v_{1i}^{(1)} - v_p \partial_\xi v_{1i}^{(3)} + v_p \partial_\zeta v_{1i}^{(3)} + v_{1i}^{(1)} \partial_\xi v_{1i}^{(1)} + v_{1i}^{(1)} \partial_\zeta v_{1i}^{(1)} + \partial_\xi \phi^{(3)} + \partial_\zeta \phi^{(3)} + v_p \partial_\zeta \varphi_{10} \partial_\xi v_{1i}^{(1)} - v_p \partial_\xi \varphi_{20} \partial_\zeta v_{1i}^{(1)}$$
$$+ \partial_\zeta \varphi_{10} \partial_\xi \phi^{(1)} + \partial_\xi \varphi_{20} \partial_\zeta \phi^{(1)} + \eta_1 \partial_{\xi\xi} v_{1i}^{(1)} + 2\eta_1 \partial_{\xi\zeta} v_{1i}^{(1)} + \eta_1 \partial_{\zeta\zeta} v_{1i}^{(1)} = 0, \tag{18}$$

$$\partial_\tau v_{2i}^{(1)} - v_p \partial_\xi v_{2i}^{(3)} + v_p \partial_\zeta v_{2i}^{(3)} + v_{2i}^{(1)} \partial_\xi v_{2i}^{(1)} + v_{2i}^{(1)} \partial_\zeta v_{2i}^{(1)} - m_{21i} \partial_\xi \phi^{(3)} - m_{21i} \partial_\zeta \phi^{(3)}$$
$$+ v_p \partial_\zeta \varphi_{10} \partial_\xi v_{2i}^{(1)} - v_p \partial_\xi \varphi_{20} \partial_\zeta v_{2i}^{(1)} - m_{21i} \partial_\zeta \varphi_{10} \partial_\xi \phi^{(1)} - m_{21i} \partial_\xi \varphi_{20} \partial_\zeta \phi^{(1)}$$
$$+ \eta_2 \partial_{\xi\xi} v_{2i}^{(1)} + 2\eta_2 \partial_{\xi\zeta} v_{2i}^{(1)} + \eta_2 \partial_{\zeta\zeta} v_{2i}^{(1)} = 0, \tag{19}$$

$$\partial_{\xi\xi} \phi^{(1)} + 2\partial_{\xi\zeta} \phi^{(1)} + \partial_{\zeta\zeta} \phi^{(1)} = n_{1i}^{(3)} - n_{2i}^{(3)} - n_{e1i} T_{e2i} \phi^{(3)} - \frac{1}{2} n_{e1i} T_{e2i}^2 \{\phi^{(1)}\}^2. \tag{20}$$

Combining Eqs. (16)-(20) and with help of equation Eq. (14) after some straight forward mathematical calculation one can obtain

$$\frac{1}{C}\left[\frac{n_{21i}}{v_p^2} v_{1i}^{(3)} - 4v_p^2 v_{2i}^{(3)}\right] = \int \left[\partial_\tau \phi_\xi^{(1)} + C_N \phi_\xi^{(1)} \partial_\xi \phi_\xi^{(1)} - C_D \partial_{\xi\xi\xi} \phi_\xi^{(1)} + C_{DP} \partial_{\xi\xi} \phi_\xi^{(1)}\right] d\zeta +$$
$$\int \left[\partial_\tau \phi_\zeta^{(1)} - C_N \phi_\zeta^{(1)} \partial_\zeta \phi_\zeta^{(1)} + C_D \partial_{\zeta\zeta\zeta} \phi_\zeta^{(1)} - C_{DP} \partial_{\zeta\zeta} \phi_\zeta^{(1)}\right] d\xi +$$
$$\iint \left[C_2 \partial_\zeta \varphi_{10} + C_3 \phi_\zeta^{(1)}\right] \partial_{\xi\xi} \phi_\xi^{(1)} d\xi d\zeta - \iint \left[C_2 \partial_\xi \varphi_{20} + C_3 \phi_\xi^{(1)}\right] \partial_{\zeta\zeta} \phi_\zeta^{(1)} d\xi d\zeta, \tag{21}$$

where

$C_N = (3 + n_{e1i} T_{e2i}^2 v_p^2 - 3n_{21i} m_{21i}^2/v_p^4)/C_1$, $C_D = C_1^{-1}$, $C_{DP} = (v_p \eta_1 + n_{21i} m_{21i} \eta_2/v_p^3)/C_1$, $C_2 = (4v_p^2 + 4n_{21i} m_{21i}/v_p^2)/C_1$, $C_3 = (n_{21i} m_{21i}^2/v_p^4 - 1 + v_p^4 n_{e1i} T_{e2i}^2)/C_1$ and $C_1 = v_p + 1 + 2n_{21i} m_{21i}/v_p^3$. $C_N$, $C_D$ and $C_{DP}$ represent the coefficient of nonlinearity, dispersion and dissipation, respectively. The first and second terms in the right side of Eq. (21) are proportional to $\zeta$ and $\xi$, respectively, because the integrand of the first term is independent of $\zeta$ and the integrand of second term is independent of $\xi$. The fourth and fifth terms in the right side of Eq. (21) are non-secular and they would be secular in the next higher order of $\varepsilon$. From Eq. (21) it is obtained

$$\partial_\tau \phi_\xi^{(1)} + C_N \phi_\xi^{(1)} \partial_\xi \phi_\xi^{(1)} - C_D \partial_{\xi\xi\xi} \phi_\xi^{(1)} + C_{DP} \partial_{\xi\xi} \phi_\xi^{(1)} = 0, \tag{22}$$
$$\partial_\tau \phi_\zeta^{(1)} - C_N \phi_\zeta^{(1)} \partial_\zeta \phi_\zeta^{(1)} + C_D \partial_{\zeta\zeta\zeta} \phi_\zeta^{(1)} - C_{DP} \partial_{\zeta\zeta} \phi_\zeta^{(1)} = 0, \tag{23}$$
$$C_2 \partial_\zeta \varphi_{10} + C_3 \phi_\zeta^{(1)} = 0, \quad C_2 \partial_\xi \varphi_{20} + C_3 \phi_\xi^{(1)} = 0. \tag{24}$$

Equations (22) and (23) represent the oppositely propagating two-sided KdVB equations in the frame of references $\xi$ and $\zeta$, respectively. It is the fact that the dissipation arises due to the cause of dissipative mechanism, such as wave-particle interaction, the effects of turbulence, dust fluctuations in dusty plasma, multi-ion streaming, Landau damping, anomalous viscosity and so on, that leads to Burger's term. In this study the Burger's term arises in Eqs. (22) and (23) due to the presence of viscosity of ion species and it is responsible for the formation of shock in the IAWs. The solution of the Eqs. (22) and (23) are obtaining as [49]

$$\phi_\xi^{(1)} = \frac{3C_{DP}^2}{25C_N C_D}\left\{sech^2\left[\frac{1}{2}\left(-\frac{C_{DP}}{5C_D}\xi + \frac{6C_{DP}^2}{125C_D^2}\tau\right)\right] + 2\left[1 - tanh\left\{\frac{1}{2}\left(-\frac{C_{DP}}{5C_D}\xi + \frac{6C_{DP}^2}{125C_D^2}\tau\right)\right\}\right]\right\}, \quad (25)$$

$$\phi_\zeta^{(1)} = \frac{3C_{DP}^2}{25C_N C_D}\left\{sech^2\left[\frac{1}{2}\left(-\frac{C_{DP}}{5C_D}\zeta - \frac{6C_{DP}^2}{125C_D^2}\tau\right)\right] + 2\left[-1 - tanh\left\{\frac{1}{2}\left(-\frac{C_{DP}}{5C_D}\zeta - \frac{6C_{DP}^2}{125C_D^2}\tau\right)\right\}\right]\right\}. \quad (26)$$

Equation (24) yields

$$\varphi_{10} = -\frac{C_3}{C_2}\int_{-\infty}^{\zeta}\phi_\zeta^{(1)}d\dot\zeta, \quad \varphi_{20} = -\frac{C_3}{C_2}\int_{+\infty}^{\xi}\phi_\xi^{(1)}d\dot\xi. \quad (27).$$

After head-on collision the phase shift of the solitons can be determine as

$$\left.\begin{array}{l}\Delta\varphi_{10} = -\varepsilon^2\dfrac{6C_3 C_{DP}^2}{25C_2 C_N C_D}\\[1em]\Delta\varphi_{20} = \varepsilon^2\dfrac{6C_3 C_{DP}^2}{25C_2 C_N C_D}\end{array}\right\}. \quad (28)$$

3.1 **Monotonic, Oscillatory and Soliton Solutions**

In the limit $C_D \to 0$, Eq. (22) reduces to

$$\partial_\tau \phi_\xi^{(1)} + C_N \phi_\xi^{(1)}\partial_\xi \phi_\xi^{(1)} + C_{DP}\partial_{\xi\xi}\phi_\xi^{(1)} = 0. \quad (29)$$

Equation (29) represents the well known Burger equation.

Let us consider,

$$\phi_\xi^{(1)}(\xi,\tau) = \phi^1(\chi), \quad \chi = \xi - V_0\tau, \quad (30)$$

where $V_0$ is the speed of the reference frame. Inserting Eq. (30) into Eq. (29) one can obtain

$$C_{DP}d_{\chi\chi}V_0 + C_N\phi^1 d_\chi\phi^1 - V_0 d_\chi\phi^1 = 0. \quad (31)$$

The solution of Eq. (31) is

$$\phi^1 = \frac{V_0}{C_N}\left\{1 - tanh\left(\frac{\xi - V_0\tau}{\frac{2C_{DP}}{V_0}}\right)\right\}. \quad (32)$$

Similarly the solution for Eq. (23) yields

$$\phi^2 = \frac{V_0}{C_N}\left\{1 + tanh\left(\frac{\xi + V_0\tau}{\frac{2C_{DP}}{V_0}}\right)\right\}, \quad (33)$$

where $\frac{V_0}{C_N}$ and $\frac{2C_{DP}}{V_0}$ represent the amplitude and width of the monotonic shock wave, respectively.

Also using Eq. (30) in Eq. (22) it is found that

$$-V_0 d_\chi \phi^1 + \frac{1}{2}C_N d_\chi \{\phi^1\}^2 + C_D d_{\chi\chi\chi}\phi^1 - C_{DP} d_{\chi\chi}\phi^1 = 0. \quad (34)$$

Integrating once, Eq. (34) yields

$$-V_0 \phi^1 + \frac{1}{2}C_N \{\phi^1\}^2 + C_D d_{\chi\chi}\phi^1 - C_{DP} d_\chi \phi^1 = 0. \quad (35)$$

Introducing $\phi^1 = \phi_1 + \phi_0$ ($\phi_1 \ll \phi_0$) into Eq. (35) and using the equilibrium condition $\phi_0 = 2V_0/C_N$ one can determine

$$C_D d_{\chi\chi}\phi_1 - C_{DP} d_\chi \phi_1 + V_0 \phi_1 = 0, \quad (36)$$

putting $\phi_1 = e^{L\chi}$ in Eq. (36) it is found that

$$C_D L^2 - C_{DP} L + V_0 = 0. \quad (37)$$

Therefore, $L = \frac{C_{DP}}{2C_D} \pm \frac{1}{2C_D}\sqrt{(C_{DP}^2 - 4C_D V_0)}$; the critical value of the dissipation coefficient is predicted as $C_{DPc} = 2\sqrt{C_D V_0}$. From this condition it is observed that the IAWs have monotonic shock for $C_{DPc} < C_{DP}$ and oscillatory shock for $C_{DPc} > C_{DP}$. The solution of Eq. (34) for oscillatory shock ($C_{DPc} > C_{DP}$) is determined as

$$\phi^1 = \frac{2V_0}{C_N} + C_{4a} e^{\frac{C_{DP}}{2C_D}(\xi - V_0\tau)} cos\{(V_0/C_D)^{1/2}(\xi - V_0\tau)\}. \quad (38)$$

Similarly, the solution for Eq. (23) can be obtained as

$$\phi^2 = \frac{2V_0}{C_N} + C_{4a} e^{\frac{C_{DP}}{2C_D}(\zeta + V_0\tau)} cos\{(V_0/C_D)^{1/2}(\zeta + V_0\tau)\}. \quad (39)$$

In special circumstances, if the dispersion dominates and the dissipation diminishes as ($C_{DP} \to 0$), then the Burger terms from KdVB equations [(22), (23)] can be neglected and under this situation Eqs. (22) and (23) are converted to the following form

$$\partial_\tau \phi_\xi^{(1)} + C_N \phi_\xi^{(1)} \partial_\xi \phi_\xi^{(1)} - C_D \partial_{\xi\xi\xi} \phi_\xi^{(1)} = 0, \quad (40)$$

$$\partial_\tau \phi_\zeta^{(1)} - C_N \phi_\zeta^{(1)} \partial_\zeta \phi_\zeta^{(1)} + C_D \partial_{\zeta\zeta\zeta} \phi_\zeta^{(1)} = 0. \quad (41)$$

Equations (40) and (41) are the well known two-sided KdV equations and they admit the soliton solutions as

$$\phi_\xi^{(1)} = \frac{3v_0}{C_N} sech^2 \left( \frac{\xi - v_0 \tau}{2\sqrt{\frac{C_D}{v_0}}} \right) \text{ and } \phi_\zeta^{(1)} = \frac{3v_0}{C_N} sech^2 \left( \frac{\xi + v_0 \tau}{2\sqrt{\frac{C_D}{v_0}}} \right). \quad (42)$$

4. **Results and Discussion**

The plasma system consisting of negative ions with positive ions and electrons are considered to study the effects of plasma parameters such as density ratios, kinematic viscosity of positive and negative ions on the head-on collision of counter propagating IASWs, formation of shock waves, phase shift after head-on collision, amplitudes as well on the soliton solution. Due to this taking into account the typical values $m_{21i} = 3.74, T_{2i} = 0.05, T_e = 0.2, 0 \leq n_{e1i} \leq 1, 0 \leq n_{21i} \leq 1$, $\eta_1 = 0.35, 0.25, 0.213, 0.188, 0.179$ and $\eta_2 = 0.001 - 0.004, 0.050, 0.1 - 0.4$.

At $C_N = 0$ the critical density ratio of positive to negative ions is equal to $n_{21i} = n_{21i(c)} = 0.64607 (\approx 0.65)$ for $T_{2i} = 0.05$, $T_e = 0.2, m_{21i} = 3.74, \eta_1 = 0.35, \eta_2 = 0.003$, and $n_{e1i} = 0.35$, which is in good agreement with the experimental [37] value ($n_{21ic} = 0.65$). Fig.1 signifies the effect of density ratio on the phase velocity of fast positive ion acoustic (PIA) mode. It is observed that the phase velocity is increasing (decreasing) with increasing of $n_{21i}$ ($n_{e1i}$). It is also seen that for $n_{e1i} > 0.45$ the phase velocity remain unchanged. The phase velocity increases with the increase of $n_{21i}$ to a point where the damping is weak, which is a significant modification of PIA mode in the limit $n_{21i} \rightarrow 1$, as $T_e \approx T_e/(1 - n_{21i})$ [12,28]. These are the conditions for the formation of a shock. This result is consistent with the experimental [37] results. Figs. 2(a) and 2(b) display the effects of $n_{21i}$ and $\eta_1$, respectively, on amplitudes. From Fig. 2(a) it is observed that the amplitudes of IASWs are increasing in the range 0 to 0.2 and then decreasing with increasing $n_{21i}$ but for $n_{e1i}$ the amplitudes of IASWs decrease. This means that the electrons contribute to the restoring force in the range $0 < n_{e1i} < 0.2$ with respect to its small number density and then increasing the restoring force in the range $n_{e1i} > 0.2$. According to the charge neutrality condition, the density of electrons decreases while the density ratio of negative to positive ions increases, causes the reduction in restoring force that provides the inertialess electrons, as a result the nonlinearity increases. For the balance between the effects of dispersion and nonlinearity, it needs to balance with the enhancement of nonlinearity of the plasma. This finally leads to the decrease in amplitudes of the IASWs. It is seen from Fig. 2(b) that the amplitudes of IASWs are decreasing due to the effect of positive ion kinematic viscosity.

Figures (3) to (6) describe the spatiotemporal evolution of electrostatic shock profiles ($\phi^{(1)}$) (i.e. the head-on collision of two IASWs) in $\xi - \zeta$ plane considering the effects of $n_{21i}$, $n_{e1i}$, $T_{2i}$, $T_e$, $\eta_1$, $\eta_2$, and $m_{21i}$. It is demonstrated in Fig. 3 and Fig. 4 that the widths of the shocks are increasing with increasing $n_{21i}$, as well as the height of the shocks are also increasing in the range $0 \leq n_{21i} < 0.65$ but the opposite result is observed in the range $0.65 < n_{21i} < 1$. Fig. 5 depicts the effects of kinematic viscosity ($\eta_2$) of the negative ion density on the formation of shock. It is observed that the widths of the shock waves decrease with the increase of $\eta_2$, maintaining the almost equal height. Figures 6 illustrate that the widths (heights) are increasing (decreasing) with the enhancement of $n_{e1i}$ in IASWs. The strength of shocks increases for the enhancement of $n_{21i}$ because the electrons gain energy due to the increase (decrease) of density of negative ions (electrons) in the concerned plasma system as shown in Figs (3) and (4). It is also observed that the compressive and rarefactive IA shocks are formed that are concerned with PIA mode depending on $n_{21i}$, $n_{e1i}$ and $\eta_2$ in the presence of negative ions. Hence, the compressive and rarefactive correspond to the compression and depression of the positive ion density that are guided by the compression and depression of the electron density, respectively. This finding is in good agreement with the experimental results [37]. The effect of the negative ion viscosity on the formation of shock can be elucidated on the basis of communal friction force between the layers of the consider plasma system. In fact the viscosity is a communal friction force between fluid layers in the considered electron-pair ion plasma. Due to increasing negative ion viscosity the frictional force increases, causes the decrease of shock width which is shown in Fig. 5. On the other hand, due to the depopulation of ion density, the density of electron increases, as a result the driving force increases that provided by the ions' inertia in IASWs.

Thus the amplitude of the shock decreases, which depicts in Fig.6. After head-on collision, if the velocity of the traveling soliton becomes slower (faster), then the phase shift would be positive (negative). Figures 7 display the effects of $n_{21i}$, $n_{e1i}$, $\eta_1$ and $\eta_2$ on phase shifts of IASWs after head-on collision. Fig. 7(a) illustrates that the phase shifts are increasing (decreasing) with increasing the density ratios $n_{21i}$ and $n_{e1i}$. While Fig.7 (b) indicates that the phase shifts are increasing with increasing $\eta_1$ but $\eta_2$ shows insignificant effect on the phase shift. It is to be noted that for both cases the phase shift is compressive, this means that the post collisional part of the trajectories lead the initial trajectories. It is to be noted that the phase shift is significantly modified due to the effects of $n_{21i}$, $n_{e1i}$ and $\eta_1$. Fig. 8 shows the post-collisional spatiotemporal potential profiles of the counter propagating IASWs. It is observed from this figure that the soliton $S_R$ travels towards right direction and the soliton $S_L$ travels towards left direction. Also the mesial line (i.e. separation line) clearly demarcates the two stages of collisions, one of it is the pre-collisional and the other is post-collisional potential profiles. This means that the propagating IASWs change their propagating plane due to collision. For the condition $C_{DPc} < C_{DP}$ ($n_{21i} > 0.311$, $T_{2i} = 0.05$, $V_0 = 0.0099$, $m_{21i} = 3.74$, $\eta_1 = 0.35$, $\eta_2 = 0.1$, $n_{e1i} = 0.6$ and $T_e = 0.2$) the monotonic shock is formed. The head-on collision of the monotonic shock solutions as a function of $n_{21i}$ are presented in Figs. 9. It is seen from these figures that the heights of the monotonic shocks are increasing with increasing $n_{21i}$ maintaining almost equal widths. Since, the dispersive coefficient becomes negligible and dissipative coefficient becomes dominant in the considered plasma system, thus the monotonic shock is formed. Figures 10 illustrate the spatiotemporal evolution of the electrostatic potential profiles of soliton solution. It is observed that both hump and dip shaped rarefective soliton exist for the effect of $\tau$. As the dispersion coefficient becomes dominant to the propagation of the concerned plasma system, thus the hump and dip shape solitons are obtained. The hump or dip shape solitary structure arises due to the balance between the effects of nonlinearity and the dispersion. For $n_{21i} < 0.311$, $T_{2i} = 0.05$, $V_0 = 0.0099$, $m_{21i} = 3.74$, $\eta_1 = 0.35$, $\eta_2 = 0.1$, $n_{e1i} = 0.6$ and $T_e = 0.2$ the condition $C_{DP} < C_{DPc}$ holds, thus the oscillatory shock is formed. Fig. 11 depicts the temporal formation of shocks of the IASWs. The amplitudes and widths of the oscillatory shocks are decreasing with increasing $\tau$ as shown in Fig.11. The negative ion density can be realized as the depopulation of positive ions in the considered plasma system, hence the driving force (provides electrons) of the IASWs decreases, consequently the amplitudes of oscillatory shocks are decreased. Figures 12 show the effect of negative ion density on the electrostatic potential profiles ($\phi^{(1)}$) along with their range of validity of the IASWs. For $n_{21i} = 0.60$ and $n_{21i} = 0.67$ it can be seen that $-0.015 \leq \phi^{(1)} \leq 0.015$ and $-0.017 \leq \phi^{(1)} \leq 0.017$ as shown in Fig.12 (a) and Fig.12 (b), respectively. Fig.12(c) depicts the potential profiles in the range $-150 \leq \phi^{(1)} \leq 150$ for $n_{21i(c)} = 0.646$, which is unexpected because the range of $\phi^{(1)}$ is $-1 \leq \phi^{(1)} \leq 1$. The positive ions density jumps into compressive shocks for $n_{21i} < n_{21i(c)}$ and negative ions density jumps into rarefactive shocks for $n_{21i} > n_{21i(c)}$. Consequently, it is remarkable to note that the threshold for the formation of shock lies in the range $-0.60 \leq n_{21i} \leq 0.67$, which is in good agreement with the experimental value of $n_{21i} = 0.65$ [37].

## 5. Conclusion

The head-on collision between two IASWs, phase shift after collision, shocks (monotonic and oscillatory) formation and soliton solution are investigated in a plasma system composing positive- and negative ions, and electrons. In order to solve the problems, two-sided KdVB equations are derived employing extended PLK method. Some important observations are summarized below:

(i) The head-on collision between two solitons obey the linear superposition condition in the consider plasma system.
(ii) The phase velocity increases with the increase of negative to positive ions density ratio, which ensures the fast PIA mode.
(iii) The amplitudes of IASWs are significantly modified by $n_{21i}$ and $\eta_1$.
(iv) The shocks produced are both compressive and rarefactive and agrees well with the experimental result.
(v) Plasma parameters significantly modify phase shift and the phase shifts become compressive in each case.
(vi) Due to the collision the counter propagating solitons change their propagation plane.
(vii) Due to formation of shock the threshold lies in the range $-0.60 \leq n_{21i} \leq 0.67$ and the result is consistent with the experimental value $n_{21i} = 0.65$.

(viii) The monotonic (oscillatory) shock must be formed in the range $n_{21i} > 0.311$ ($n_{21i} < 0.311$) for particular values of the plasma parameters.

(ix) Both the hump and dip shape rarefactive soliton are existed in the soliton solution.

The results obtained might be useful to understand the shock formation experiment in the laboratory as well as in the space plasma consisting of positive- and negative ions, and electrons. It may also be useful to understand the features of broadband shock noise in the D- and F-regions of the Earth's ionosphere.

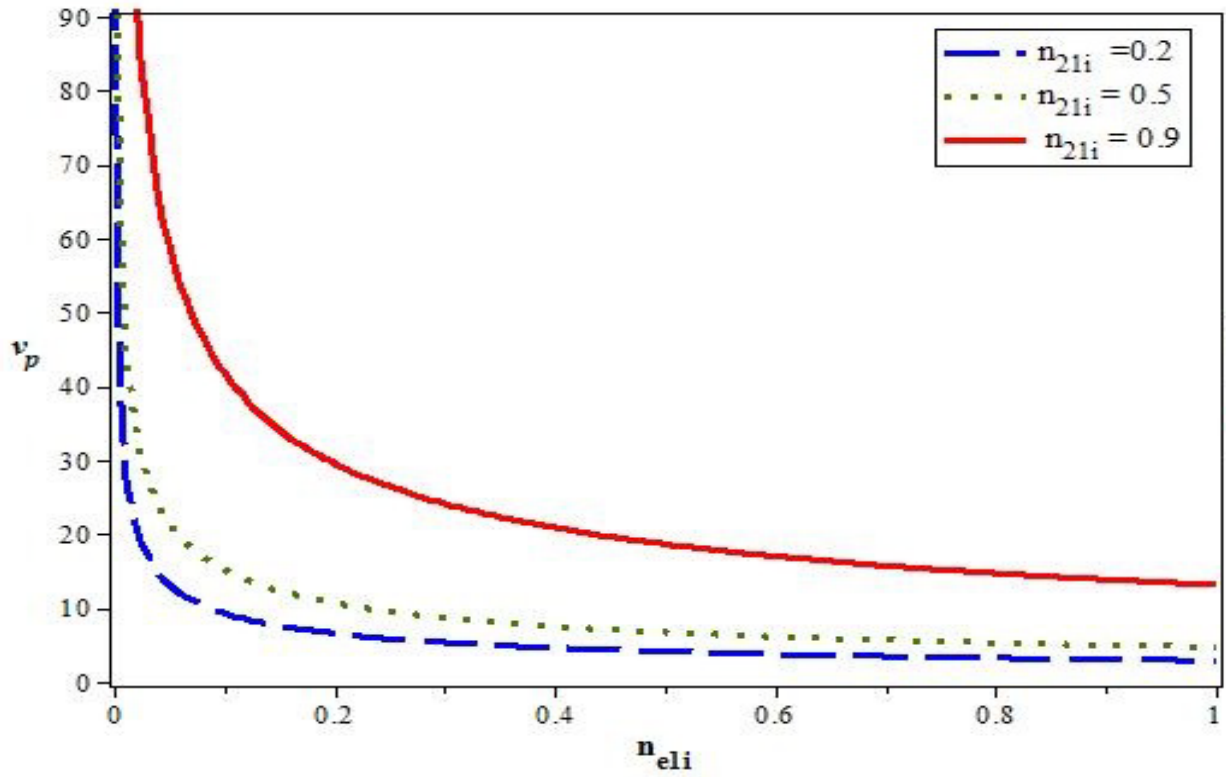

**Fig.1** Effects of $n_{21i}$ on $v_p$ of PIA mode taking $v_0 = 0.005$, $m_{21i} = 3.74$, $\eta_1 = 0.30$, $\eta_2 = 0.05$, $T_{2i} = 0.05$ and $T_e = 0.2$.

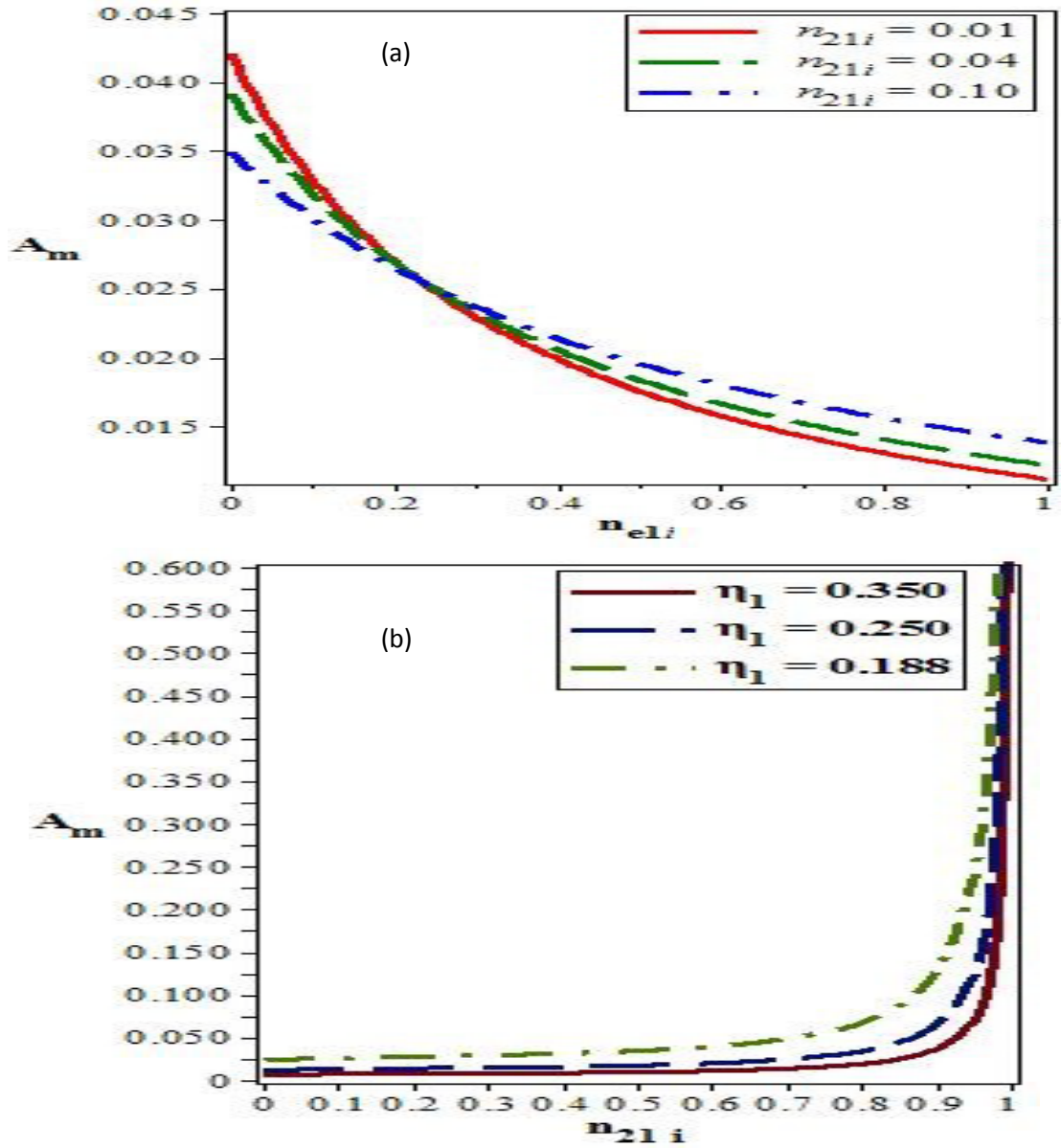

Fig.2 The amplitudes of IASWs for various values of (a) $n_{21i} = 0.01, 0.04, 0.10$ taking $v_0 = 0.005$, $m_{21i} = 3.74$, $\eta_1 = 0.30$, $\eta_2 = 0.05$, $T_{2i} = 0.05$, $T_e = 0.2$, and $T_{2ie} = T_e/(1 - n_{21i})T_{2i}$ and (b) $\eta_1 = 0.188, 0.250, 0.350$, taking $n_{21i} = 0.1$, $n_{e1i} = 0.5$ and other parameters similar to (a).

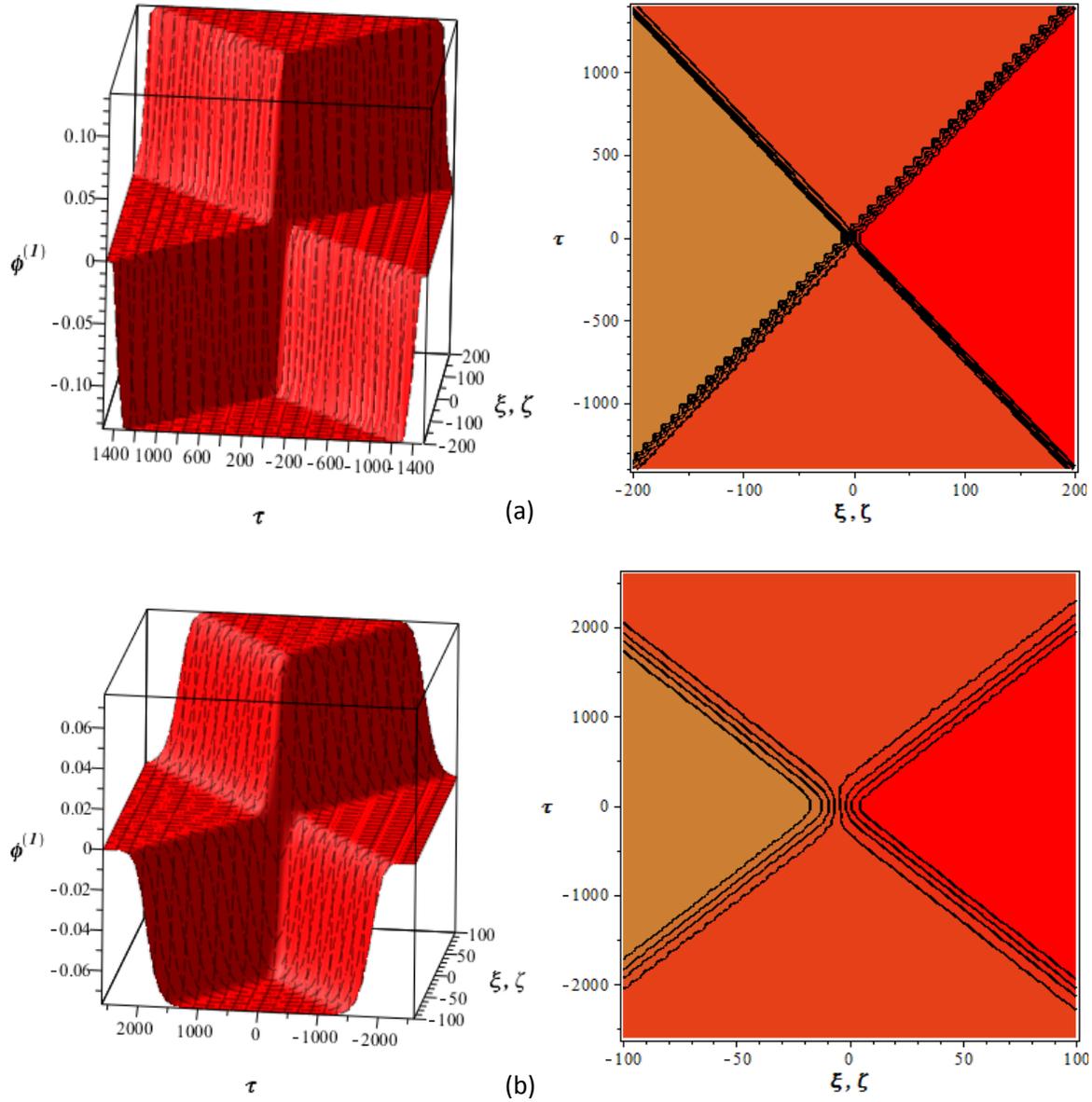

Fig.3 Electrostatic potential profile ($\phi^{(1)}$) of IASWs for (a) $n_{21i} = 0.54$ (b) $n_{21i} = 0.04$ taking $n_{e1i} = 0.35$, $T_{2i} = 0.05$, $T_e = 0.20$, $\eta_1 = 0.35$, $\eta_2 = 0.003$, $m_{21i} = 3.74$ and $T_{2ie} = T_e/(1 - n_{21i})T_{2i}$. Right column is the contour plot of the left column.

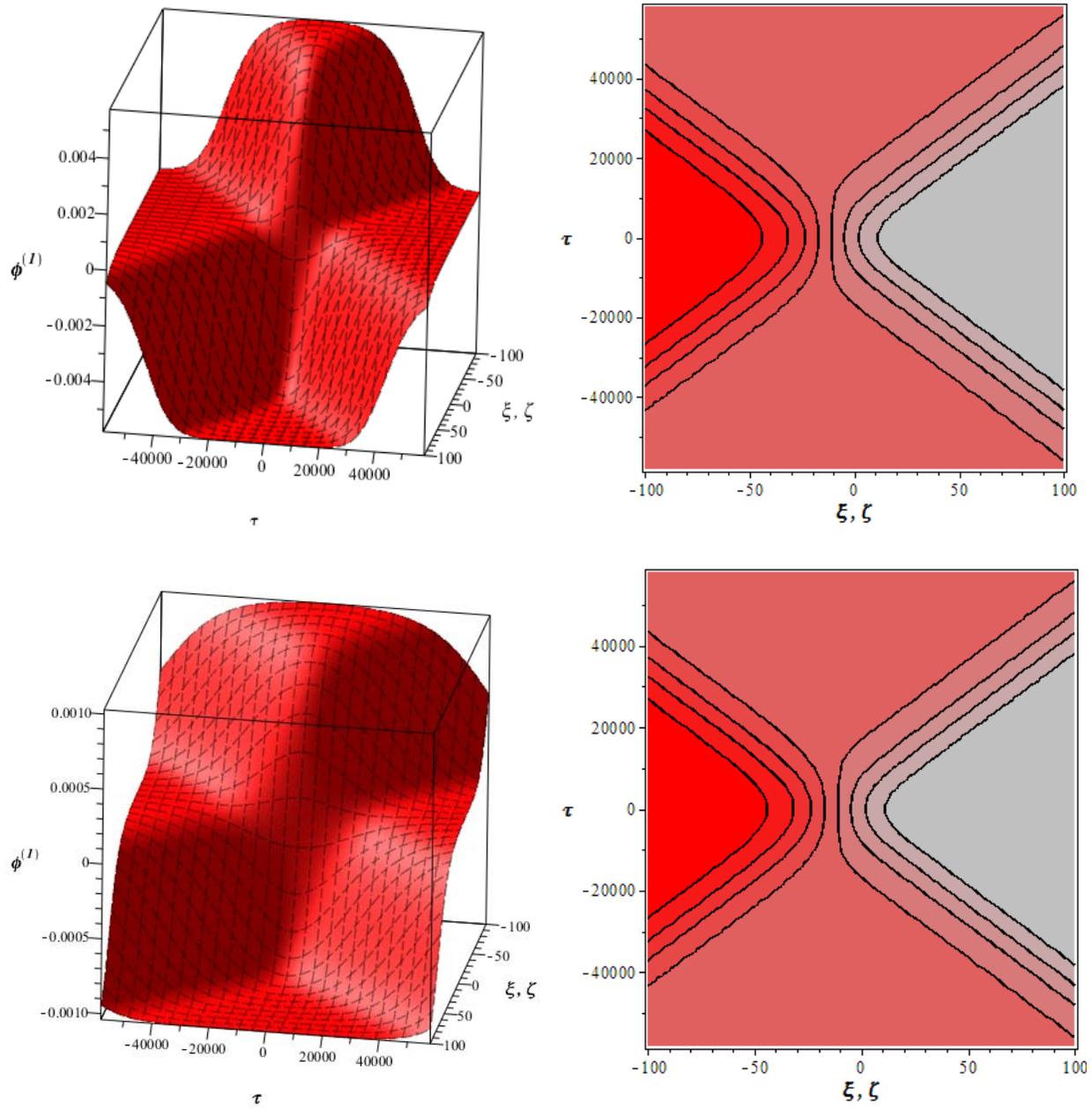

Fig.4 Electrostatic potential profile ($\phi^{(1)}$) of IASWs for (a) $n_{21i} = 0.70$ and (b) $n_{21i} = 0.78$ taking $\eta_2 = 0.003, T_{2i} = 0.05, T_e = 0.20, \eta_1 = 0.35$ $m_{21i} = 3.74$ $n_{e1i} = 0.35$ and $T_{2ie} = T_e/(1 - n_{21i})T_{2i}$. Right column is the contour plot of the left column.

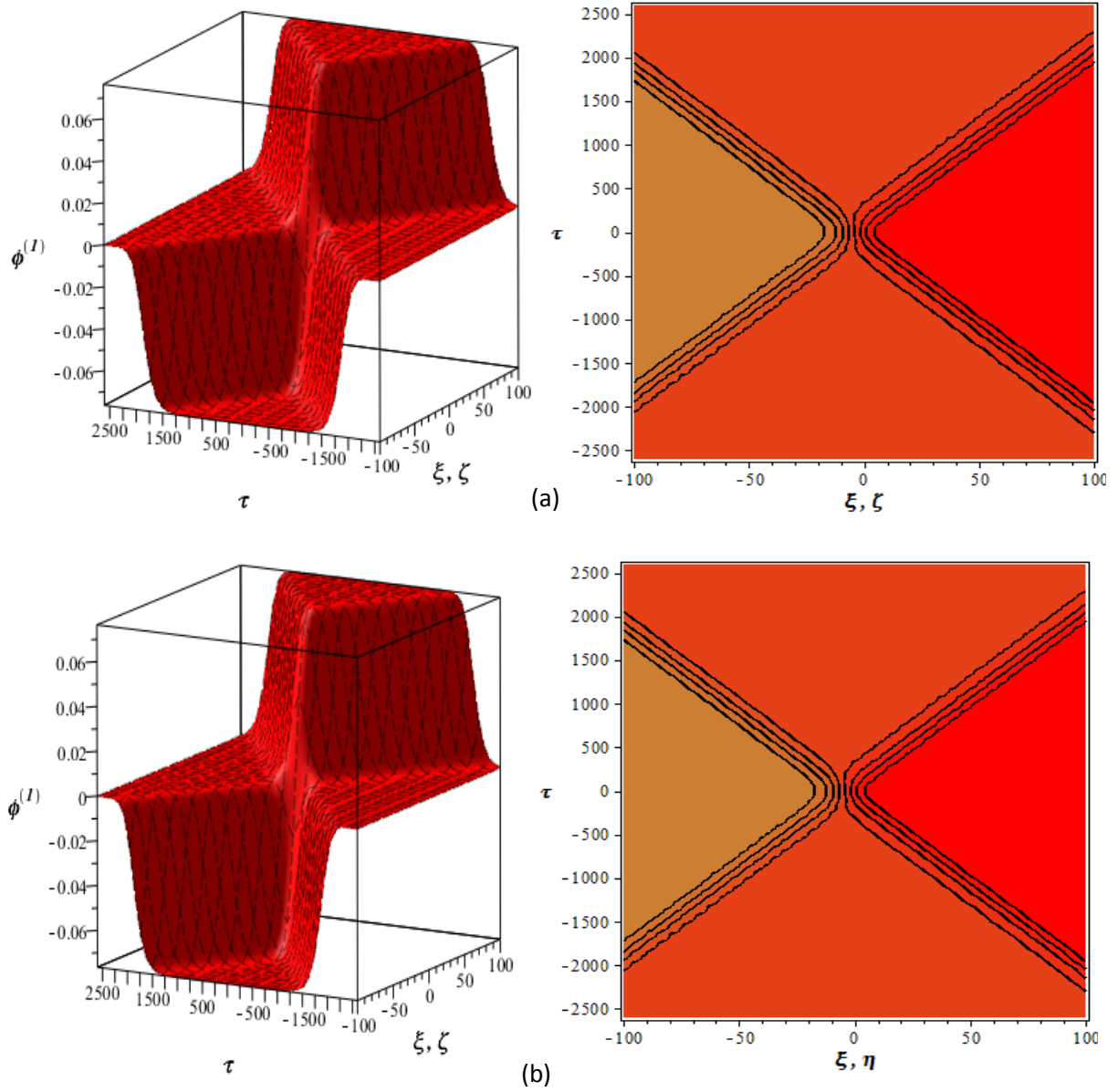

Fig.5 Electrostatic potential profile ($\phi^{(1)}$) of IASWs for (a) $\eta_2 = 0.001$ and (b) $\eta_2 = 0.003$ taking $n_{e1i} = 0.35$, $n_{21i} = 0.04$, $T_{2i} = 0.05$, $T_e = 0.20$, $\eta_1 = 0.35$, $m_{21i} = 3.74$ and $T_{2ie} = T_e/(1 - n_{21i})T_{2i}$. Right column is the contour plot of the left one.

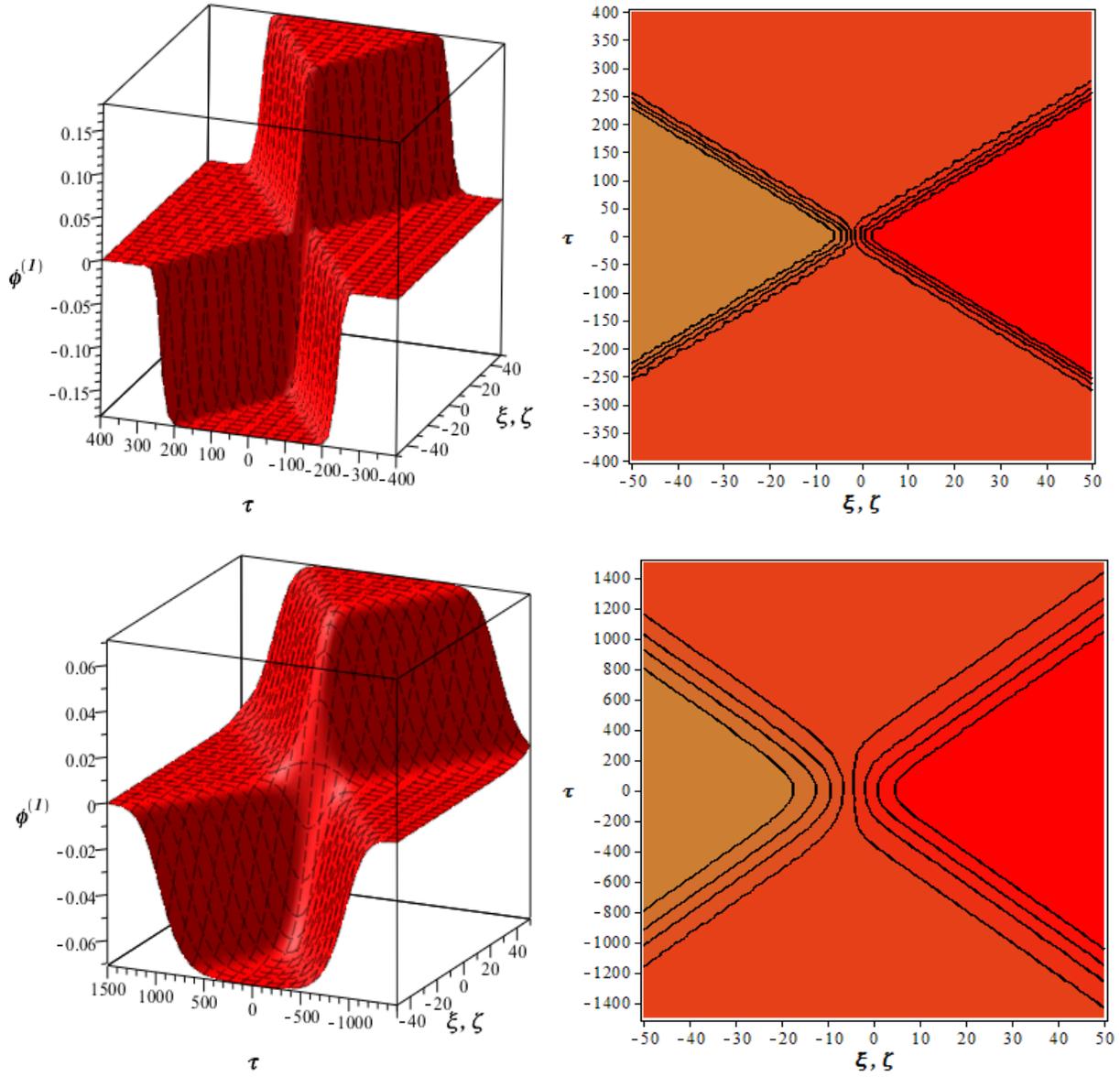

Fig.6 Electrostatic potential profiles ($\phi^{(1)}$) of IASWs for (a) $n_{e1i} = 0.35$ and (b) $n_{e1i} = 0.90$ taking $\eta_2 = 0.003$, $T_{2i} = 0.05$, $T_e = 0.20$, $\eta_1 = 0.35$  $m_{21i} = 3.74$ and $T_{2ie} = T_e/(1 - n_{21i})T_{2i}$. Right column is the contour plot of the left column.

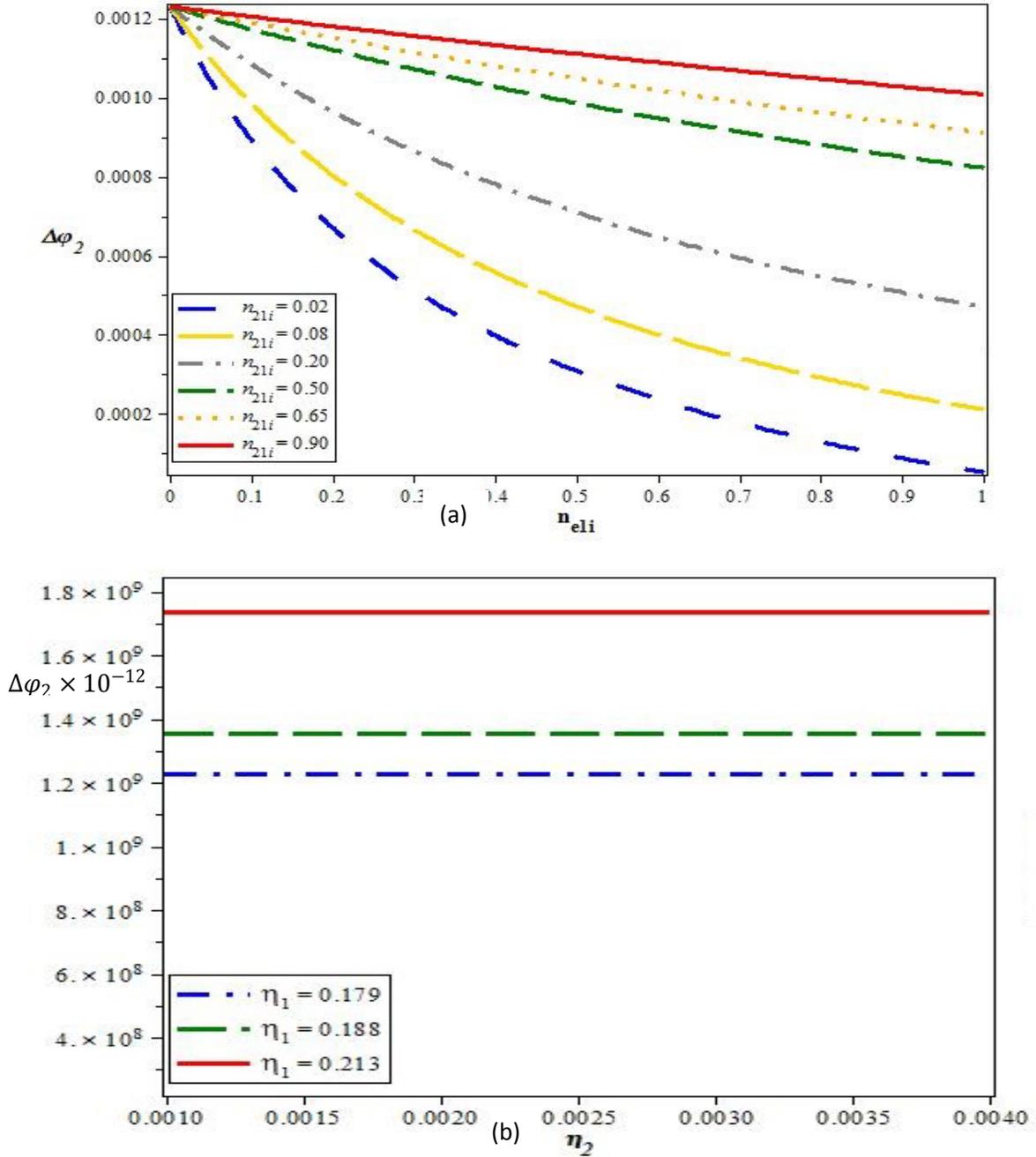

Fig.7 Changes of $\Delta\varphi_2$ of IASWs for (a) $n_{21i} = 0.02, 0.08, 0.20, 0.50, 0.65, 0.90$ taking $\eta_1 = 0.35$, $\eta_2 = 0.004$, $T_{2i} = 0.05$, $T_e = 0.20$, $m_{21i} = 3.74$, $n_{e1i} = 0.80$ and (b) $\eta_1 = 0.179$, $0.188$, $0.213$ taking $\eta_2 = 0.004$, $T_{2i} = 0.05$, $T_e = 0.20$, $n_{21i} = 0.04$, $m_{21i} = 3.74$, $n_{e1i} = 0.80$ and $T_{2ie} = T_e/(1 - n_{21i})T_{2i}$

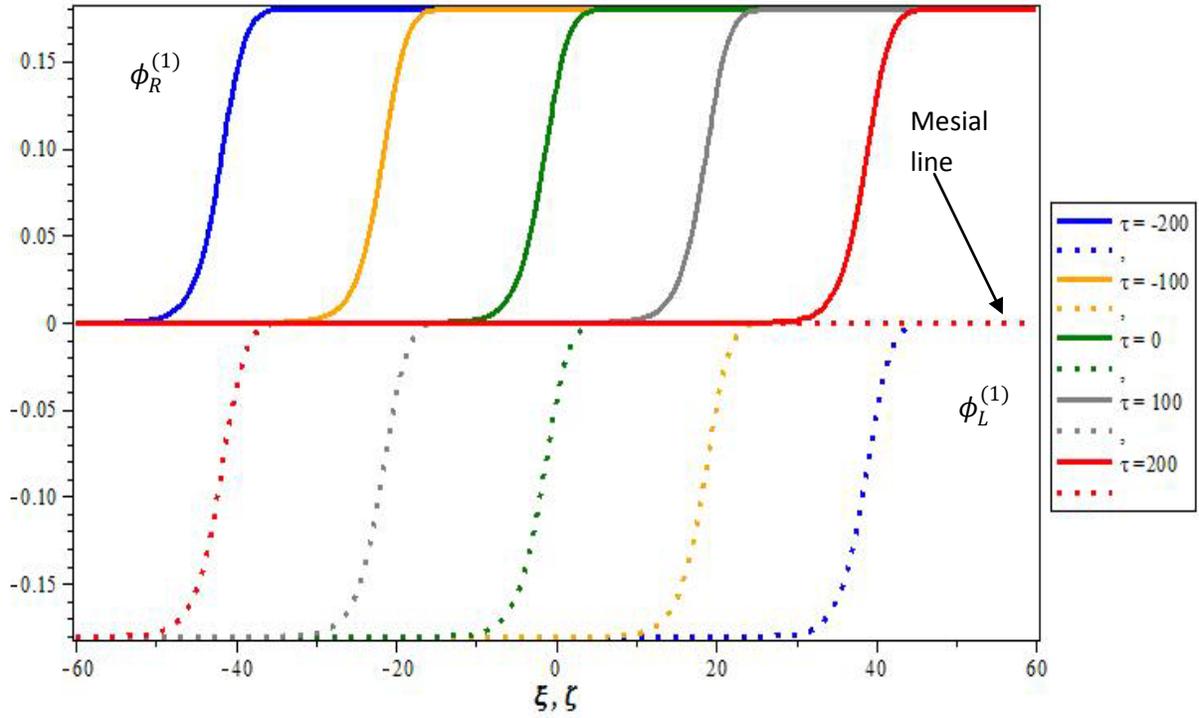

Fig. 8 Propagating process of IASWs for right (R) and left (L) traveling solitons after head-on collision, taking $\eta_1 = 0.35$, $\eta_2 = 0.004$, $T_{2i} = 0.05$, $T_e = 0.20$, $n_{21i} = 0.04$, $m_{21i} = 3.74$, $n_{e1i} = 0.80$ and $T_{2ie} = T_e/(1 - n_{21i})T_{2i}$.

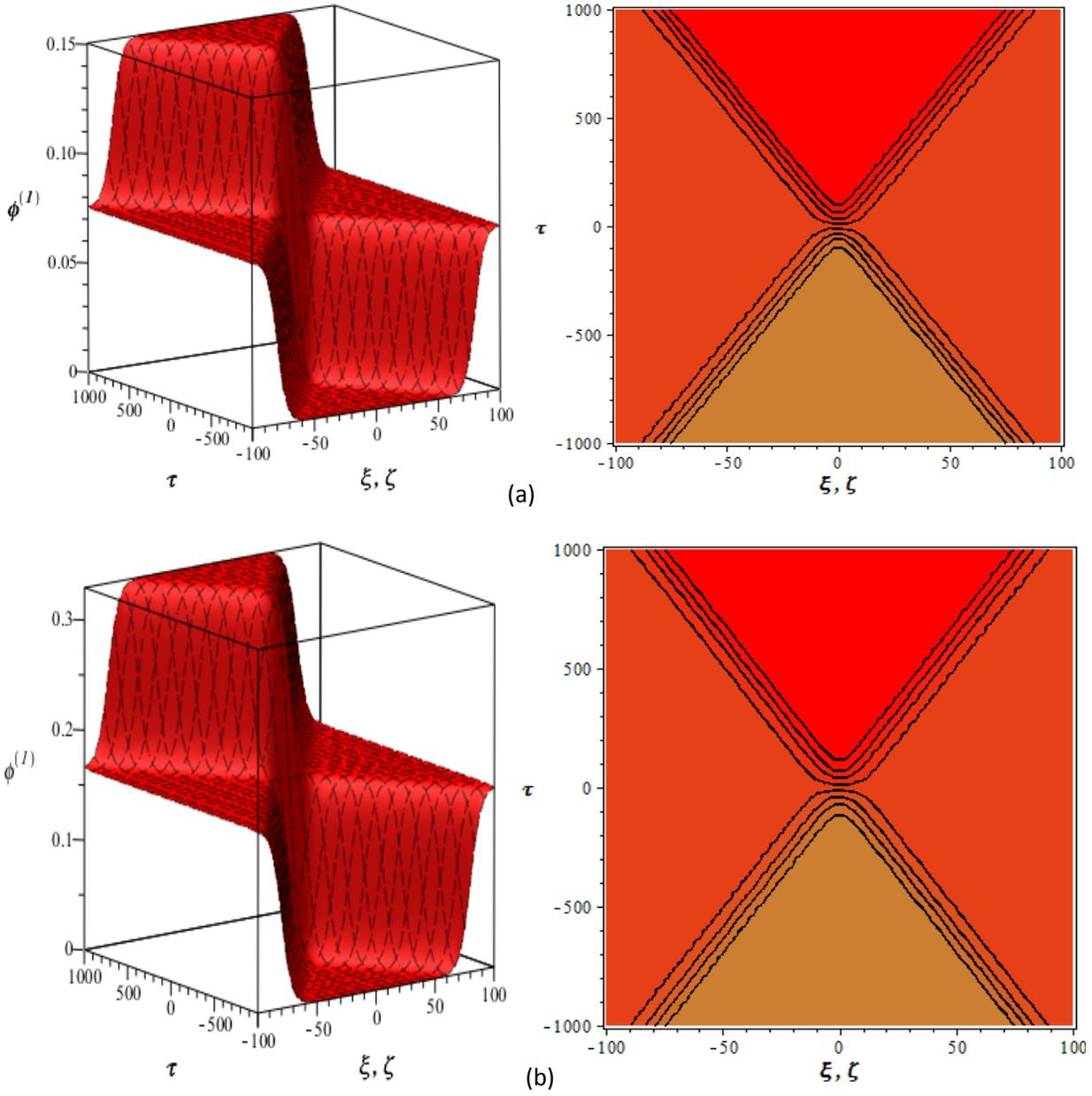

Fig.9 Monotonic shock formation $[\phi^{(1)}]$ for (a) $n_{21i} = 0.40$ and (b) $n_{21i} = 0.98$ taking $T_{2i} = 0.05$, $V_0 = 0.0099$, $m_{21i} = 3.74$, $\eta_1 = 0.35$, $\eta_2 = 0.10$, $n_{e1i} = 0.60$, $T_e = 0.20$ and $T_{2ie} = T_e/(1 - n_{21i})T_{2i}$. Right column is the contour plot of the left column.

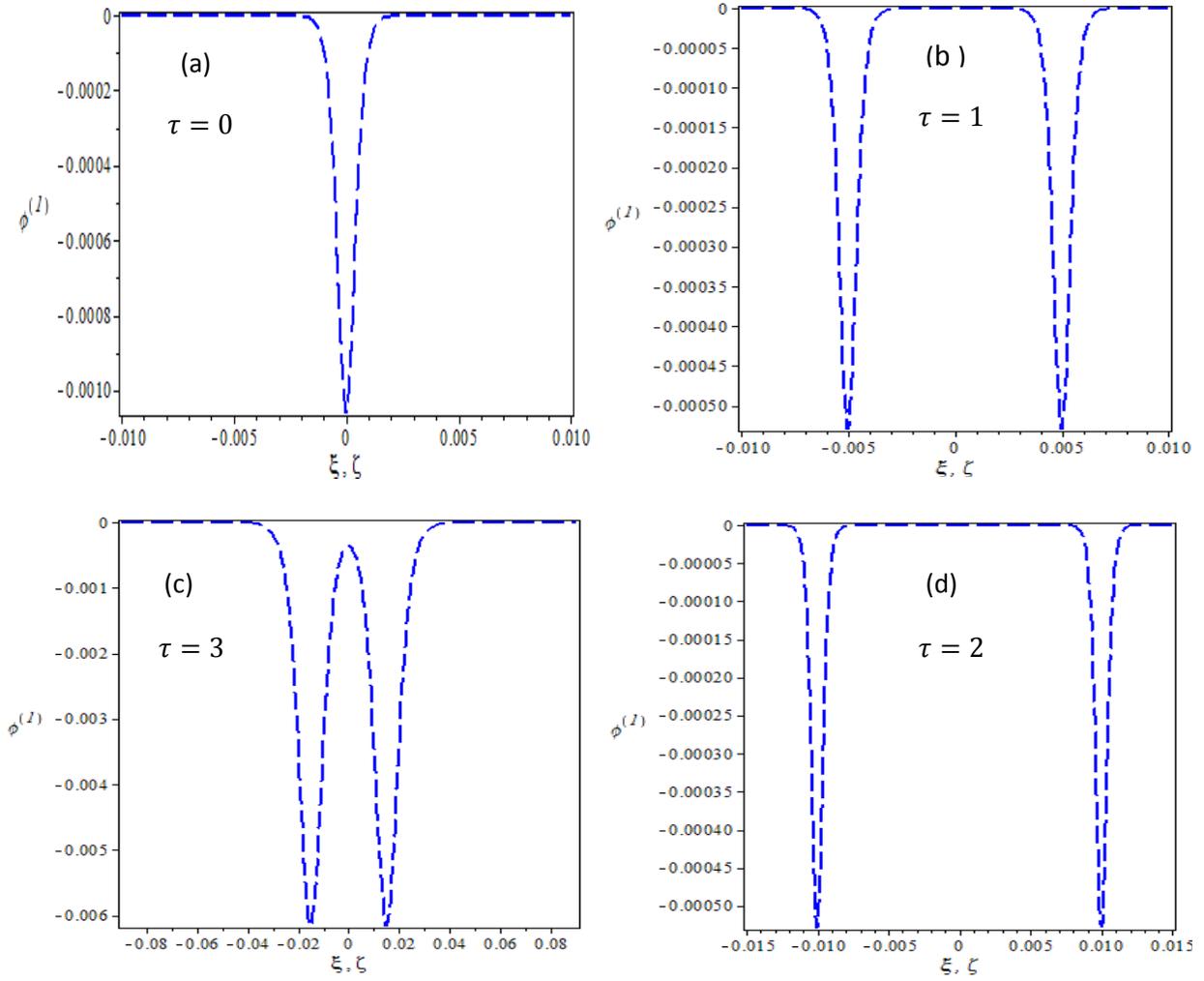

Fig. 10 Effect on electrostatic potential [$\phi^{(1)}$] of soliton solution for (a) $\tau = 0$, $n_{21i} = 0.98$ $T_{2i} = 0.05$, $V_0 = 0.005$, $m_{21i} = 3.74$, $\eta_1 = 0.35$, $\eta_2 = 0.3$, $n_{e1i} = 0.60$, $T_e = 0.2$, $T_{2ie} = T_e/(1 - n_{21i})T_{2i}$, (b) $\tau = 1$ and other parameters same as (a), (c) $\tau = 3$, $n_{21i} = 0.09$ and other parameters same as (a), and (d) $\tau = 2$ and other parameters same as (a).

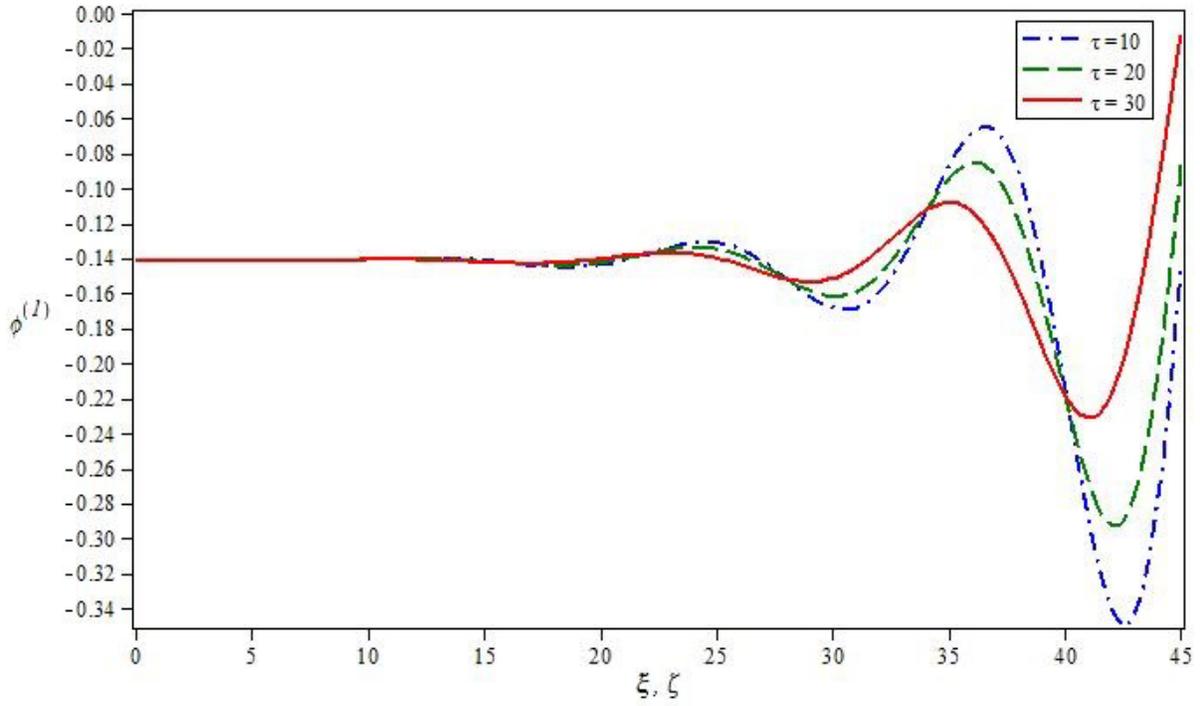

Fig.11 Formation of oscillatory shocks [$\phi^{(1)}$] for $\tau = 10, 20, 30$ taking $n_{21i} = 0.08$ $T_{2i} = 0.05$, $v_0 = 0.0099$, $m_{21i} = 3.74$, $\eta_1 = 0.35$, $\eta_2 = 0.10$, $n_{e1i} = 0.60$, $T_e = 0.20$, $C_{4a} = 0.0001$ and $T_{2ie} = T_e/(1 - n_{21i}) T_{2i}$.

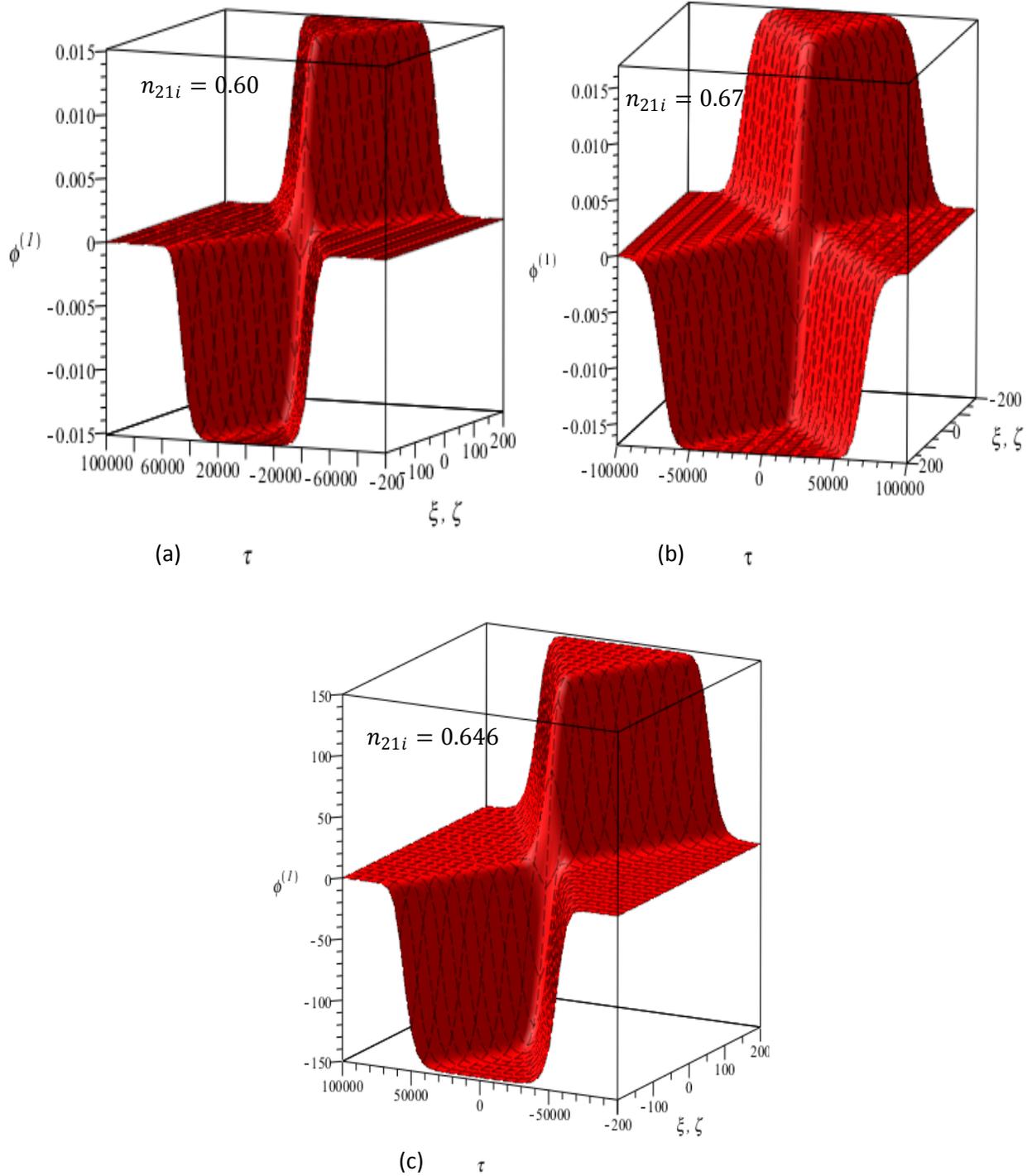

Fig.12 Electrostatic potential profiles ($\phi^{(1)}$) of IASWs for (a) $n_{21i} = 0.60$, (b) $n_{21i} = 0.67$ and (c) $n_{21i} = 0.646$, taking $n_{e1i} = 0.35$, $T_{2i} = 0.05$, $T_e = 0.20$, $\eta_1 = 0.35$, $m_{21i} = 3.74$, $n_{21i} = 0.04$, $\eta_2 = 0.001$ and $T_{2ie} = T_e/(1 - n_{21i}) T_{2i}$